\documentclass[11pt,a4paper]{article}
\pdfoutput=1
\usepackage{jcappub}
\usepackage{graphicx}

\title{
Study of the performance of a large scale water-Cherenkov detector (MEMPHYS)
}

\author[a]{Luca Agostino,}
\author[a]{Margherita Buizza-Avanzini,}
\author[b]{Marcos Dracos,}
\author[c]{Dominique Duchesneau,}
\author[a,0]{Michela Marafini\note{Currently at Universit\`a La Sapienza, Rome, Italy},}
\author[d]{Mauro Mezzetto,}
\author[e]{Luigi Mosca,}
\author[a]{Thomas Patzak,}
\author[a]{Alessandra Tonazzo}
\author[a,b]{and Nikolaos Vassilopoulos}

\affiliation[a]{ APC, Univ. Paris Diderot, CNRS/IN2P3, CEA/Irfu, Obs. de Paris, Sorbonne Paris Cit\'e, F-75205 Paris Cedex 13, France }
\affiliation[b]{ IPHC, Universit\'e de Strasbourg, CNRS/IN2P3, F-67037 Strasbourg, France}
\affiliation[c]{ LAPP, Universit\'e de Savoie, CNRS/IN2P3, F-74941 Annecy-le-Vieux, France}
\affiliation[d]{ INFN Sezione di Padova, I-35131 Padova, Italy}
\affiliation[e]{ Laboratoire Souterrain de Modane, F-73500 Modane, France}

\emailAdd{luca.agostino@apc.univ-paris7.fr, buizza@in2p3.fr, marcos.dracos@ires.in2p3.fr, duchesneau@lapp.in2p3.fr, michela.marafini@roma1.infn.it, mauro.mezzetto@pd.infn.it, luigi.mosca@lsm.in2p3.fr, thomas.patzak@apc.univ-paris7.fr, tonazzo@in2p3.fr, vassilo@ires.in2p3.fr}

\abstract{
MEMPHYS (MEgaton Mass PHYSics)
is a proposed large-scale water Cherenkov experiment to be performed
deep underground. It is dedicated to nucleon decay searches, neutrinos from
supernovae, solar and atmospheric neutrinos, as well as neutrinos from
a future Super-Beam or Beta-Beam to measure 
the CP violating phase 
in the leptonic sector and the mass hierarchy. 
A full simulation of the detector has been performed to
evaluate its performance for beam physics.
The results are given in terms of ``migration matrices''
of reconstructed versus true neutrino energy,
taking into account all the experimental effects.
}

\arxivnumber{1206.6665}

\begin{document}

\maketitle

\flushbottom

\section{Introduction}

A megaton-scale water Cherenkov detector 
would have competitive capabilities 
for accelerator-based
neutrino oscillation physics.
In addition, it would reach a sensitivity on the proton 
lifetime close to the predictions of most supersymmetric or higher 
dimension grand unified theories
and it would explore neutrinos from supernovae and from other
astrophysical sources. 

Such a detector is most attractive because it relies on 
a well established technique, already used by the IMB \cite{IMB}, 
KamiokaNDE \cite{kam} 
and SuperKamiokande \cite{superk} (SK) experiments. 
Each tank will be roughly 10 times the size of SK,
a reasonable extension of a known, well performing detector.

An expression of interest for such a project, called MEMPHYS
(MEgaton Mass PHYSics), was prepared~\cite{loi}.

The potential for neutrino physics with 
specific Super-Beams and Beta-Beams at the Fr\'ejus site was investigated in detail 
in~\cite{jec}. 
The authors assumed the same performance 
as the SK detector in terms of detection efficiency,
particle identification capabilities and background rejection.
The behaviour of a larger scale detector will, however, be different,
because of the larger distance traveled by light to reach the
photomultipliers.

In this paper,  a realistic evaluation of the expected MEMPHYS performance
is presented. It is
based on a full simulation and complete reconstruction and analysis algorithms.
``Migration matrices'' from true to reconstructed neutrino energy
are provided.

\section{The MEMPHYS detector}
\label{sec-Detector}

MEMPHYS is a proposed large-scale water Cherenkov detector
with a fiducial mass of the order of half a megaton.

\begin{figure}[t]
\begin{center}
\begin{tabular}{cc}
\includegraphics[width=0.3\linewidth]{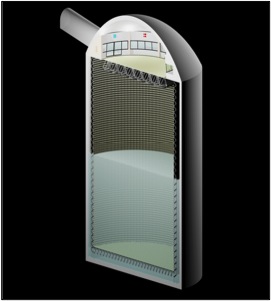} &
\includegraphics[width=0.6\linewidth]{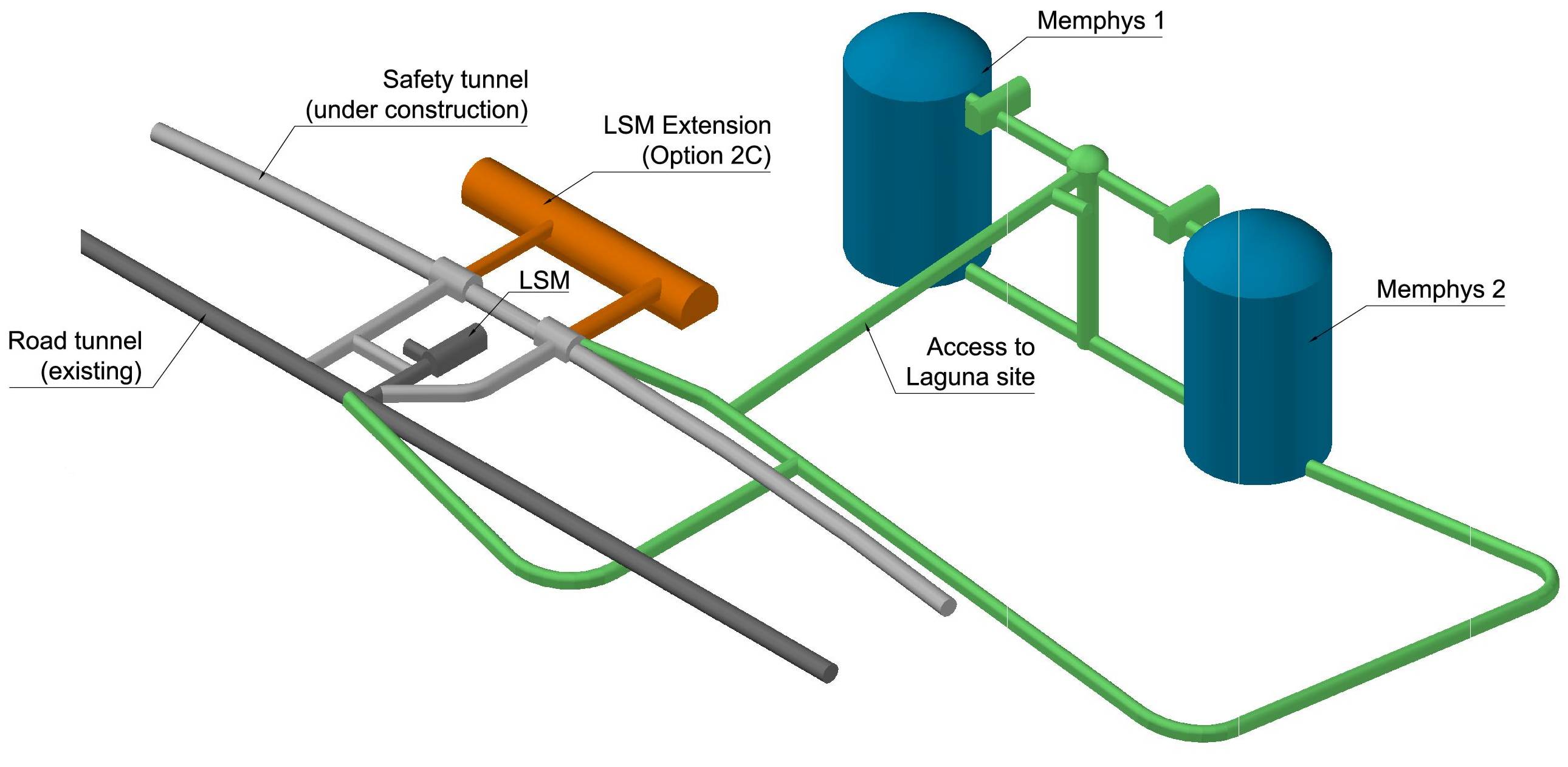}
\end{tabular}
\end{center}
\caption{Schematic view of one MEMPHYS module (left) and
design for installation and infrastructure at 
a possible extension of the LSM underground laboratory
at the Fr\'ejus site (right).
Each tank is 65 m in diameter and 103 m in height. The total 
fiducial mass is 500 kton.
\label{fig:fig1}
}
\end{figure}

The detector
could be installed at the Fr\'ejus site, near the existing 
{\em Laboratoire Souterrain de Modane} (LSM
laboratory), in the tunnel connecting France to
Italy, located at 130 km from CERN and with a rock overburden of 4800 m.w.e.
Possible installation at other European sites was studied in the context 
of the LAGUNA EU-FP7 Design Study~\cite{laguna}.

The original plan~\cite{loi} envisaged 3 cylindrical
detector modules of 65 meters in diameter and 60 meters in height. 
At the Fr\'ejus
site, the characteristics of the rock allow for a larger 
excavation in the vertical direction.
Heights up to 103 m are possible, which would allow for the same total 
fiducial mass with only two modules.
The latest design~\cite{nimpatzak} 
envisages 2 modules of 103 m height and 65 m diameter.
Taking into account a 1.5 m thick veto volume surrounding the main tank
and a cut at 2 m from the inner tank wall for the definition of the fiducial volume,
as done in SK to allow for Cherenkov cone development,
the total fiducial mass would be 500 kilotons.

Each module is equipped with $\sim$120000 8'' or 10'' photomultipliers (PMTs) providing
30\% optical coverage (equivalent, in terms of number of collected 
photoelectrons, to the
40\% coverage with 20'' PMTs of SK).

A schematic view of the detector and of 
a possible layout for installation at the Fr\'ejus site
are shown in figure~\ref{fig:fig1}.\footnote{Courtesy of Lombardi Engineering S.A.}

\section{MEMPHYS simulation}  

In order to evaluate realistic performance for the above-described
baseline detector, a detailed simulation has been developed,
mainly in the context of the EUROnu FP7 Design Study~\cite{euronu}.
The code, based on the Geant-4 toolkit~\cite{geant4-1,geant4-2}, 
was originally written for the T2K-2km detector~\cite{mfechner},
then interfaced with the OpenScientist framework~\cite{gbarrand}.
It allows for interactive event viewing, batch processing and 
analysis.
Special care has been devoted to the modularity of the code
in the definition of the detector geometry, to facilitate future
detector optimisation studies.
The GENIE~\cite{genie} event generator is used for neutrino interactions.

The model implemented in Geant-4
for light propagation in water includes the effects of
Rayleygh scattering, Compton scattering and absorption.
The attenuation length 
%as a function of wavelength is shown in Figure~\ref{fig:attlen}. 
%It 
thus obtained
is very similar to the one shown by the SK
Collaboration~\cite{sksolar}, providing evidence for the
reliability of the simulation.

\begin{figure}[htb]
\center{
%\begin{tabular}{ccc}
\includegraphics[width=0.45\linewidth]{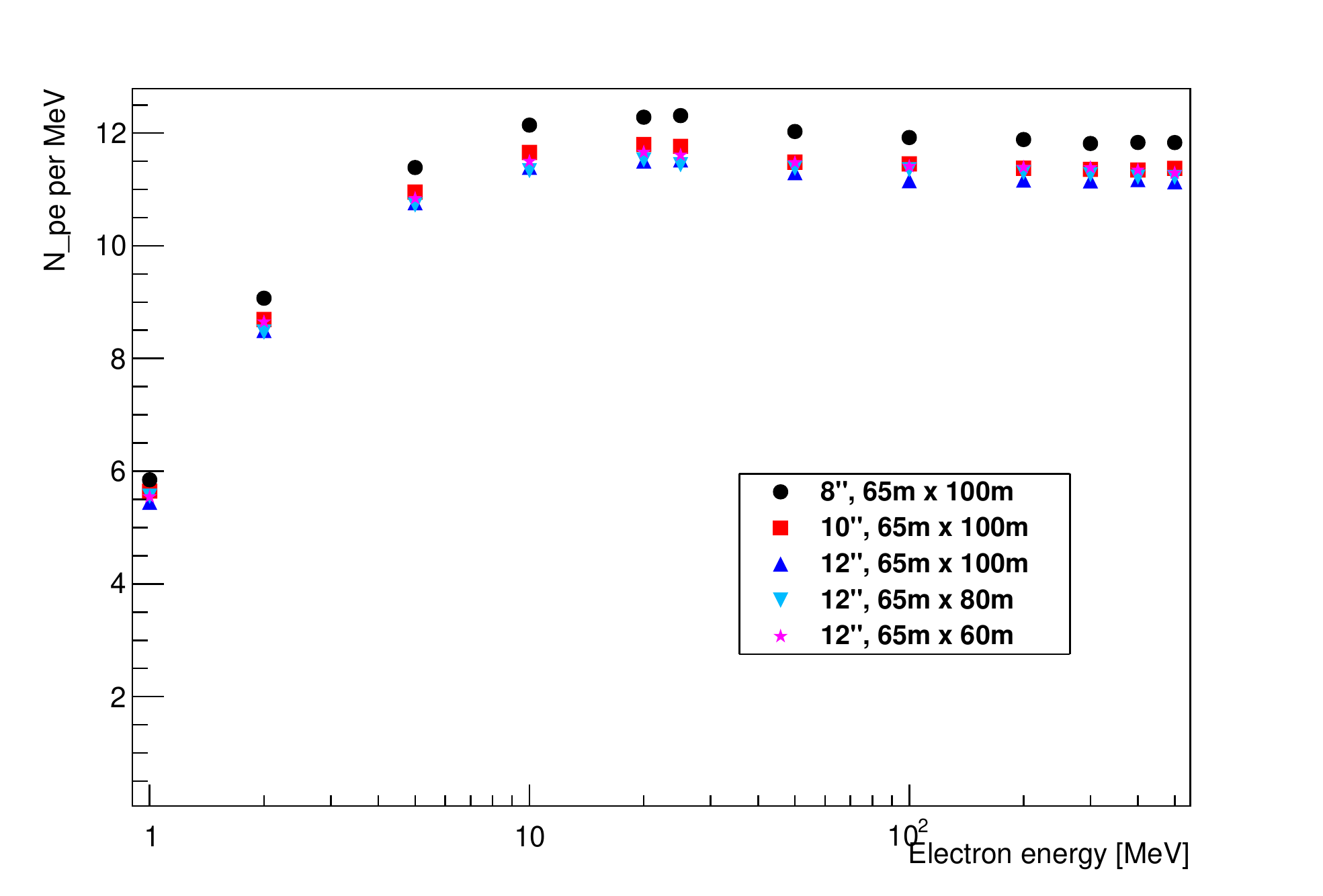} %&
\includegraphics[width=0.45\linewidth]{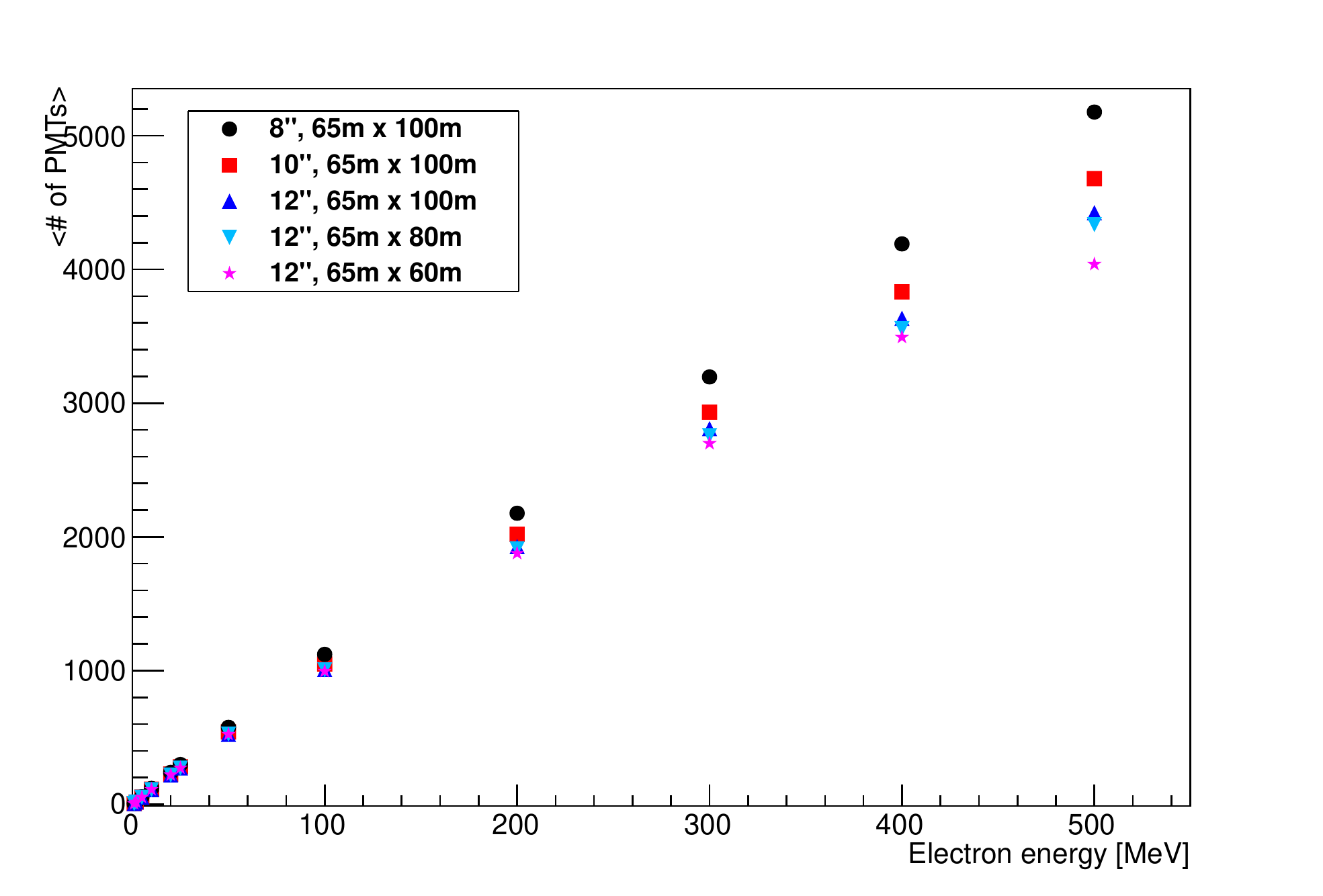} %&
\includegraphics[width=0.45\linewidth]{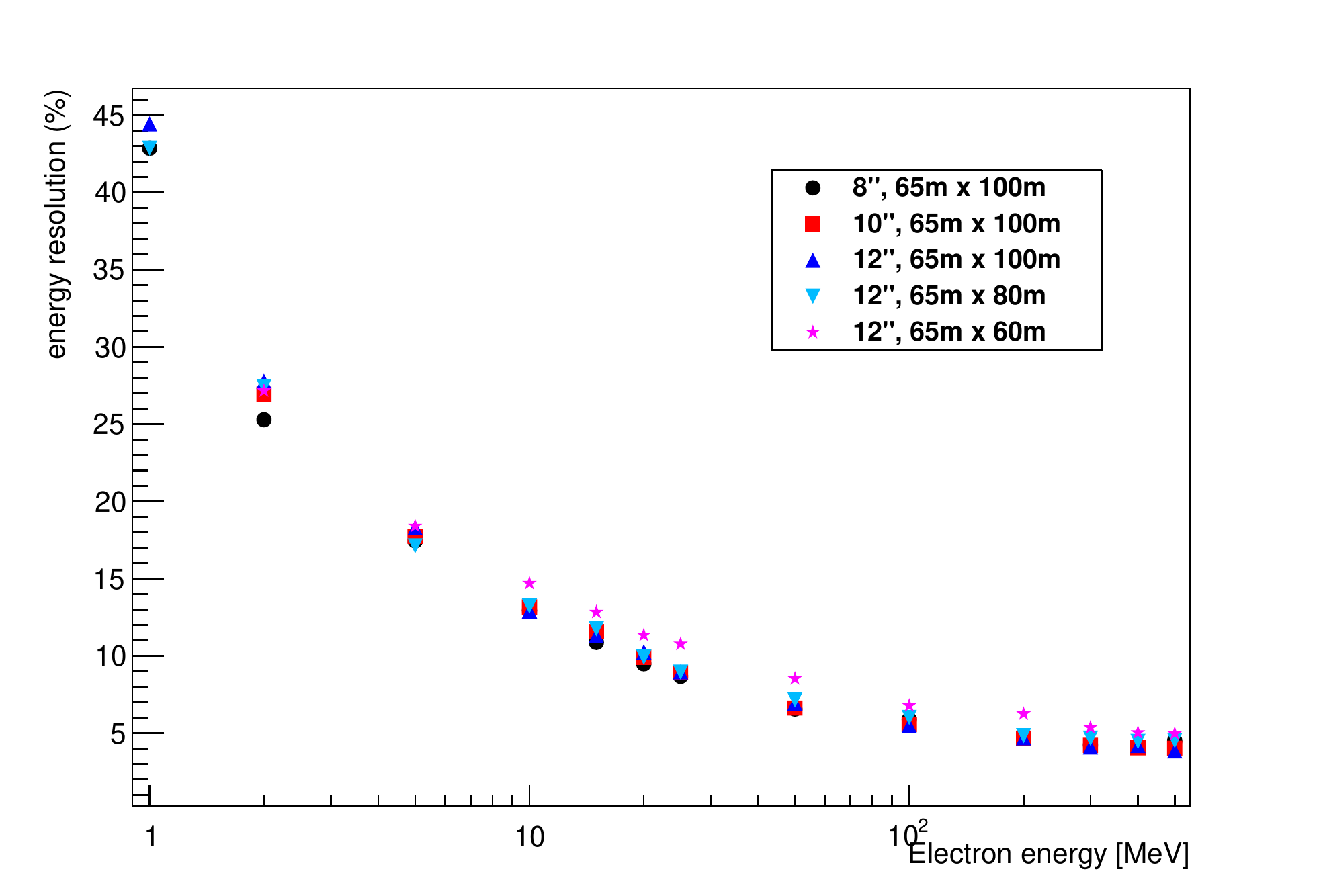}% \\
%\end{tabular}
\caption{
Detector response for different tank heights (60, 80 and 100m) 
and PMT sizes (8'', 10'' and 12'').
Top left: Number of detected photoelectrons per MeV as a function 
of electron energy.
Top right: Number of PMTs with at least one photoelectron as a function
of electron energy.
Bottom: Momentum resolution as a function
of electron energy.
\label{fig:npes}
}
}
\end{figure}

One basic quantity used to evaluate the detector performance is 
the number of photoelectrons (PEs) per MeV
as a function of the particle energy.
This is shown in figure~\ref{fig:npes}(a),
for electrons generated uniformly in the detector volume.
The number of PEs per MeV 
is about constant and equal to 11 for energies above 5 MeV.
Figure~\ref{fig:npes}(b) shows the number of hit PMTs as a function of energy.
The resolution on the estimated electron momentum, 
evaluated as explained below, is shown in Figure~\ref{fig:npes}(c).

As can be seen from the figures,
these quantities are very weakly affected by the tank height,
thus we can conclude that this parameter has no significant impact on the detector's performance.
The new baseline configuration, with 
2 tanks of 65 m diameter and 103 m height, is used in the following.

The response is also nearly unaffected by PMT size.
For the simulation presented here, 12'' PMTs were actually used: they provide identical response 
to 10'' PMTs and only slight differences with respect to 8'' PMTs.

\begin{figure}[t]
\center{
\includegraphics[width=0.45\linewidth]{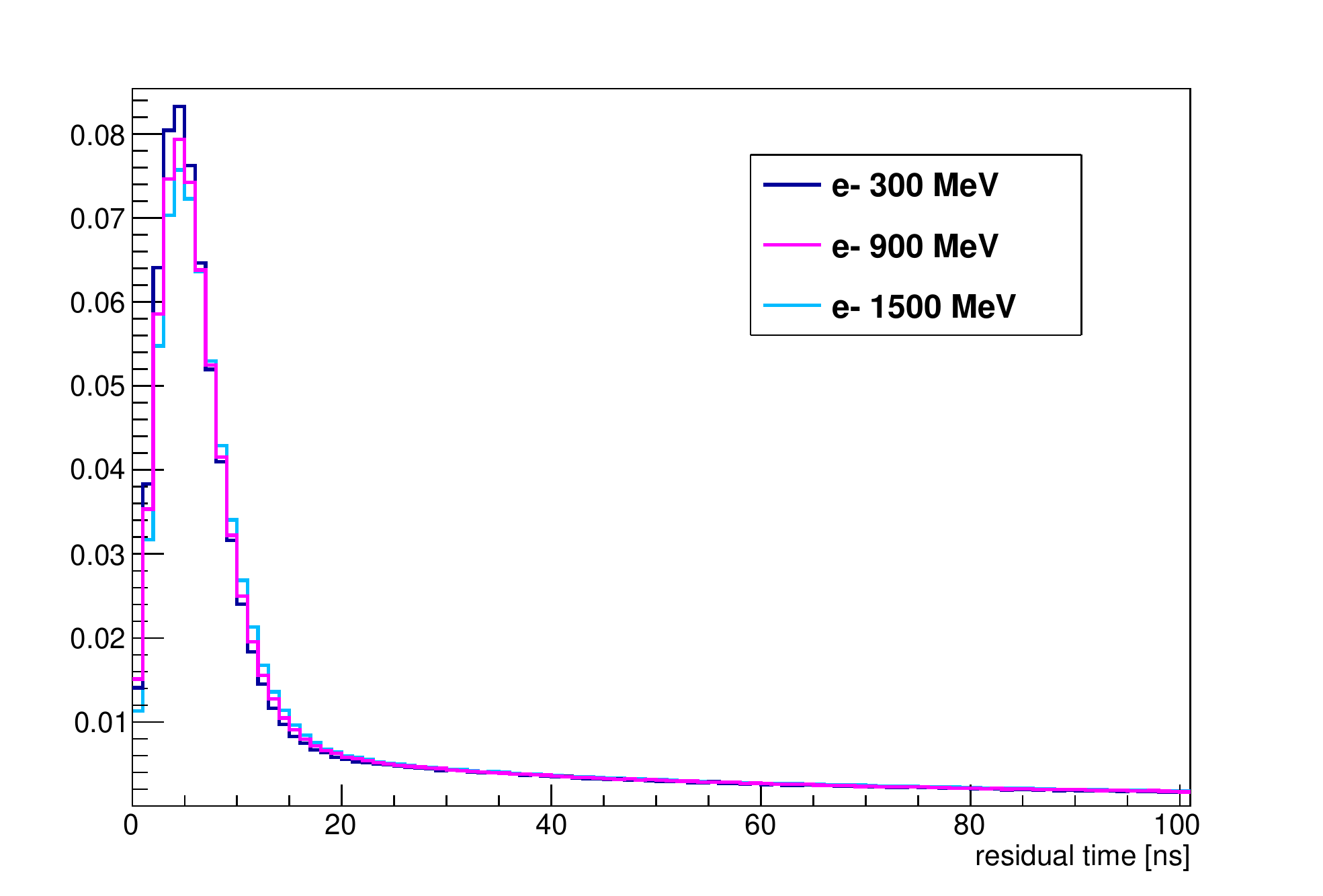}
\includegraphics[width=0.45\linewidth]{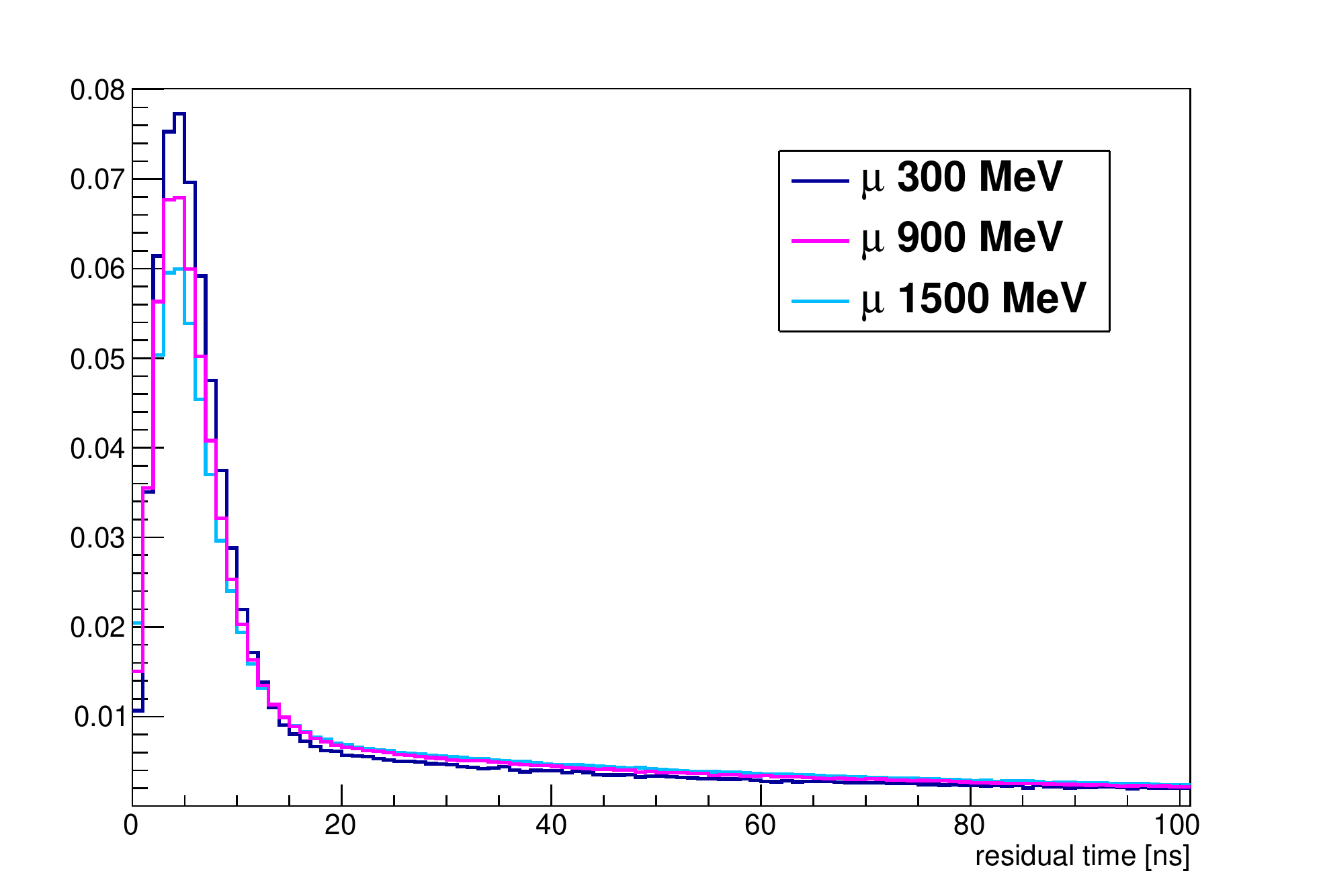}     
\caption{
Distribution of the PMTs residual times with respect to the reconstructed vertex,
for electrons (left) and muons (right) of different energies.
\label{fig:tresid}
}
}
\end{figure}

\begin{figure}[t]
\center{
\includegraphics[width=0.9\linewidth]{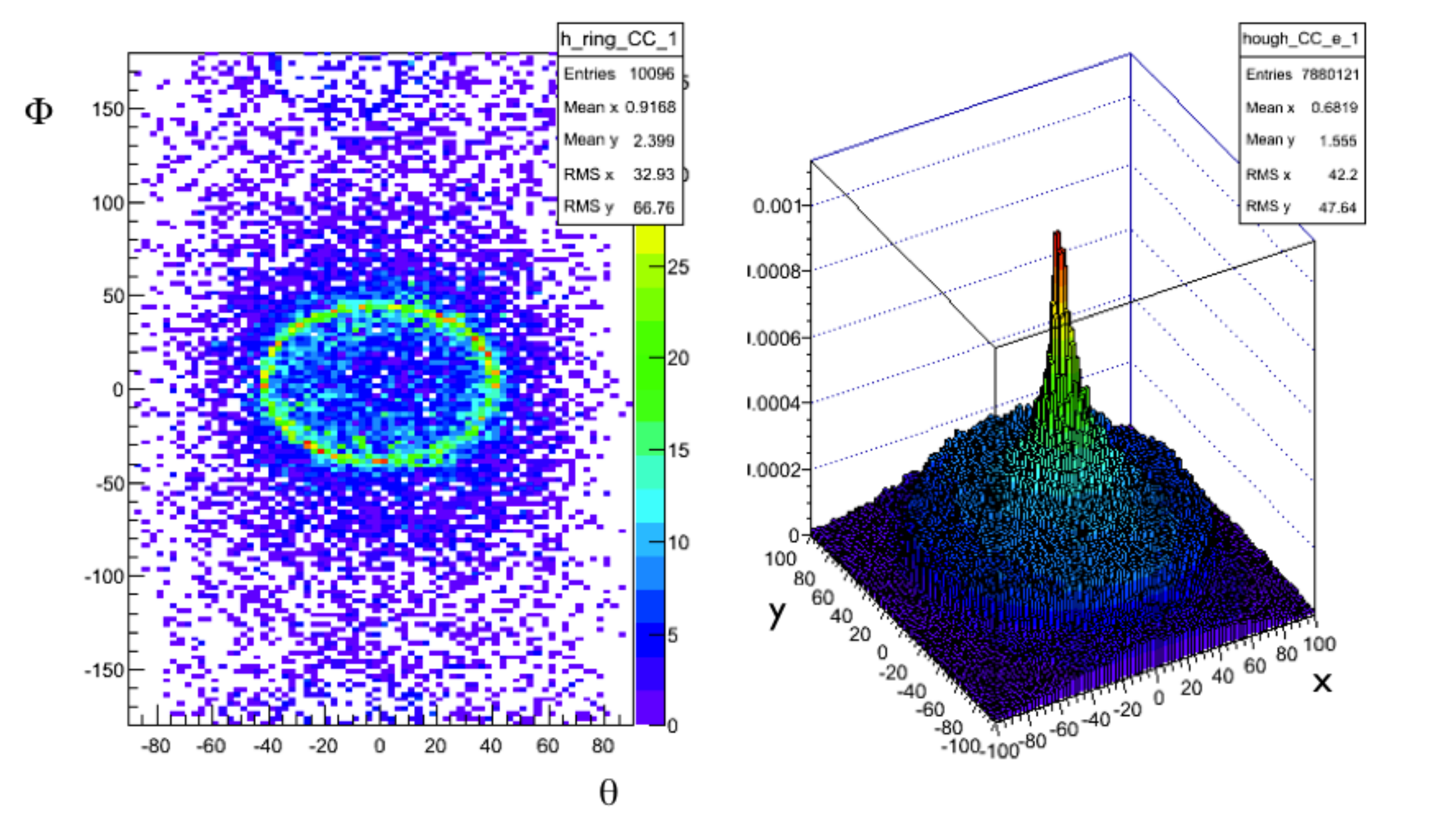} \\   
\includegraphics[width=0.9\linewidth]{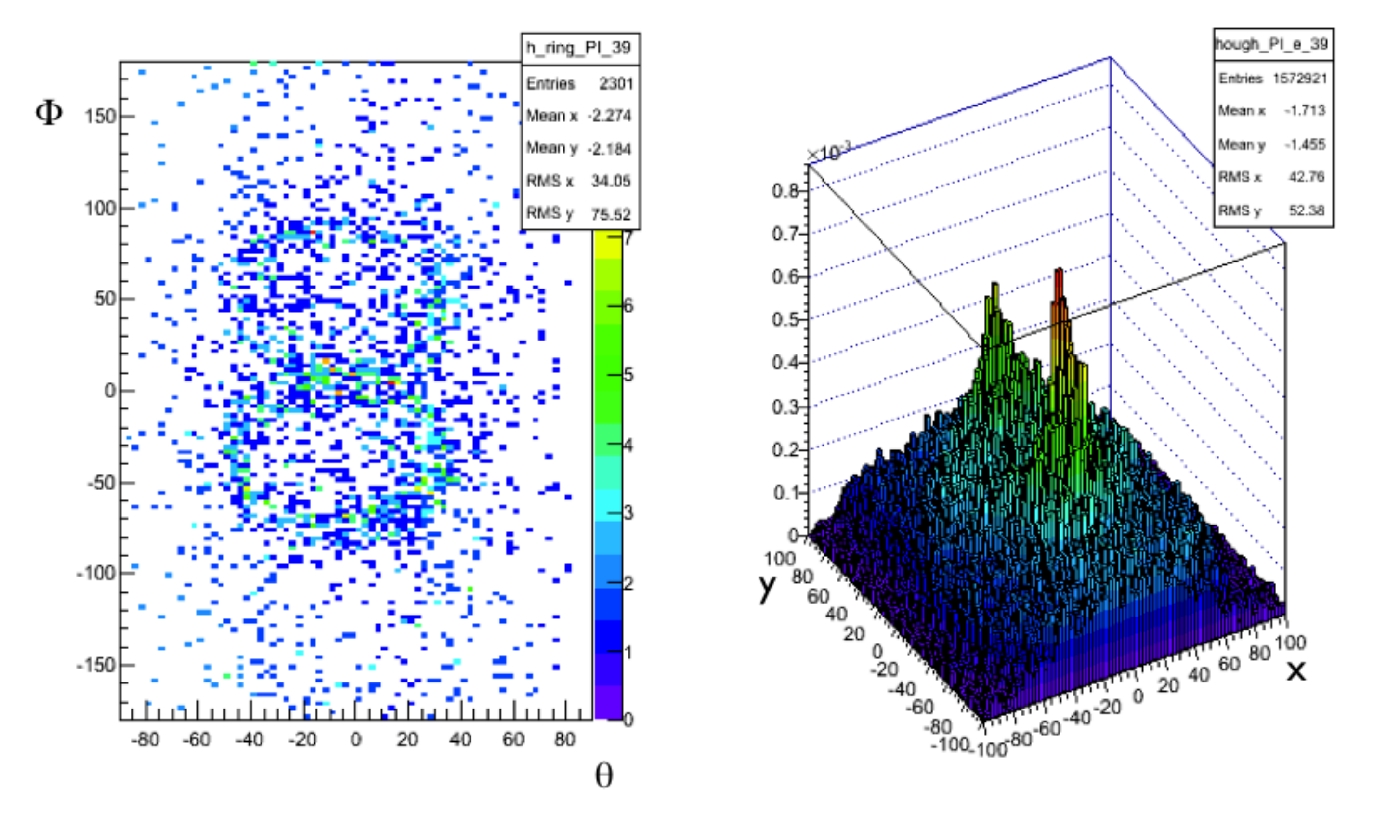}     
\caption{
Single ring (top) and double ring (bottom) events: projection
in spherical coordinates centered on the reconstructed vertex and direction
(left) and their Hough transform (right).
\label{fig:rings}
}
}
\end{figure}

The impact of PMT noise on
beam neutrino physics is not expected
to be a major issue. With a dark count rate of 7 kHz per PMT
and an event hit integration time of 50 ns, the number of spurious hits
per event will be of the order of 40, much smaller than our threshold 
on the minimum number of hits for event reconstruction (500 hits, corresponding to about
50 MeV of reconstructed electron momentum).
The use of a 50 ns integration window, as in SK, is justified by the distribution 
of the PMT hit time residuals with respect to the reconstructed vertex, shown in
Figure~\ref{fig:tresid}: the large majority of the hits lie within this range.
PMT noise was however implemented in the simulation, and its impact was evaluated on 
some features of the analysis, as explained below.

\section{Event selection and energy reconstruction}

A complete analysis chain has been developed, based on what is done
in SK~\cite{shiozawa}. Some of the algorithms
are a simplified version of those of SK.
Their performance was also evaluated by running the full simulation
with the SK parameters (size, PMT coverage etc.)
to ensure that no significant degradation of efficiencies and background rejection
are introduced by our algorithms. A rescaling was then applied to account for the
small differences due to our simplifications.

The aim of the %event reconstruction and analysis procedure 
analysis is 
the reconstruction of the incoming neutrino energy and 
the identification of its flavour, to perform appearance or disappearance measurements
with different types of beams. This is only relevant for Charged Current (CC) 
neutrino interactions. Neutral Current (NC) interactions where a final-state pion 
can mimic an electron or muon are considered separately.

The analysis proceeds through the following steps:
\begin{itemize}
\item reconstruction of the interaction vertex, from the timing of the hits in the different PMTs;
\item determination of the outgoing lepton direction, from the pattern of the Cherenkov ring;
\item lepton identification, from the ``fuzziness'' of the Cherenkov ring: 
since electrons are more subject to bremsstrahlung and multiple scattering, 
they produce rings whose edge is less ``sharp'' than in those of muons.
A simplified particle identification algorithm is used, considering the fraction of
charge inside the edge of the ring;
\item rejection of NC interaction with a $\pi^0$ in the final state, which can mimic an electron 
(more details are given below);
\item reconstruction of the lepton momentum, from the measured charge in the PMTs;
\item finally, determination of the incoming neutrino energy.
\end{itemize}

One of the most severe backgrounds in the search for $\nu_e$ appearance is due to 
NC events with a $\pi^0$ in the final state: the two $\gamma$'s originating from its decay
produce rings similar to those of electrons, and the rejection of these events
is mainly based on the reconstruction of a second ring in an electron-like event.
Figure~\ref{fig:rings} shows examples of a single-ring (electron)
and a double-ring ($\pi^0$) event:
the rings are first projected in spherical coordinates centered on the fitted
particle vertex and direction, then Hough-transformed~\cite{hough} to peaks for automated counting.
The $\pi^0$ identification algorithm used in this analysis is much simplified 
with respect to the one used in SK and in the HyperKamiokande LOI~\cite{hyperk}:
in particular, we don't implement a cut on the invariant mass of two rings, when
a second ring is forced to be found.
We have applied our $\pi^0$ identification algorithm 
selection on a sample of neutrinos interacting in a detector simulated
with an approximate SK geometry (40 m diameter, 40 m height, 40\% optical
coverage with 20'' PMTs), and we have found a $\pi^0$ contamination very similar to 
what we find with the MEMPHYS simulation, namely 3.9\%: this
suggests that we can assume the efficiency of the selection to be
nearly independent of the detector size.
The efficiencies of the $\pi^0$ rejection cut were rescaled to those
of~\cite{hyperk}, considering that we will eventually implement their full
likelihood analysis and cuts. 
A cut on the Michel electron from muon decay was also implemented;
this cut introduces some differences between muon neutrino and anti-neutrino identification
efficiency,
and in addition suppresses completely the $\nu_e$ contamination in the $\nu_\mu$ sample.

\begin{figure}[t]
\center{
\begin{tabular}{cc}
\includegraphics[width=0.5\textwidth]{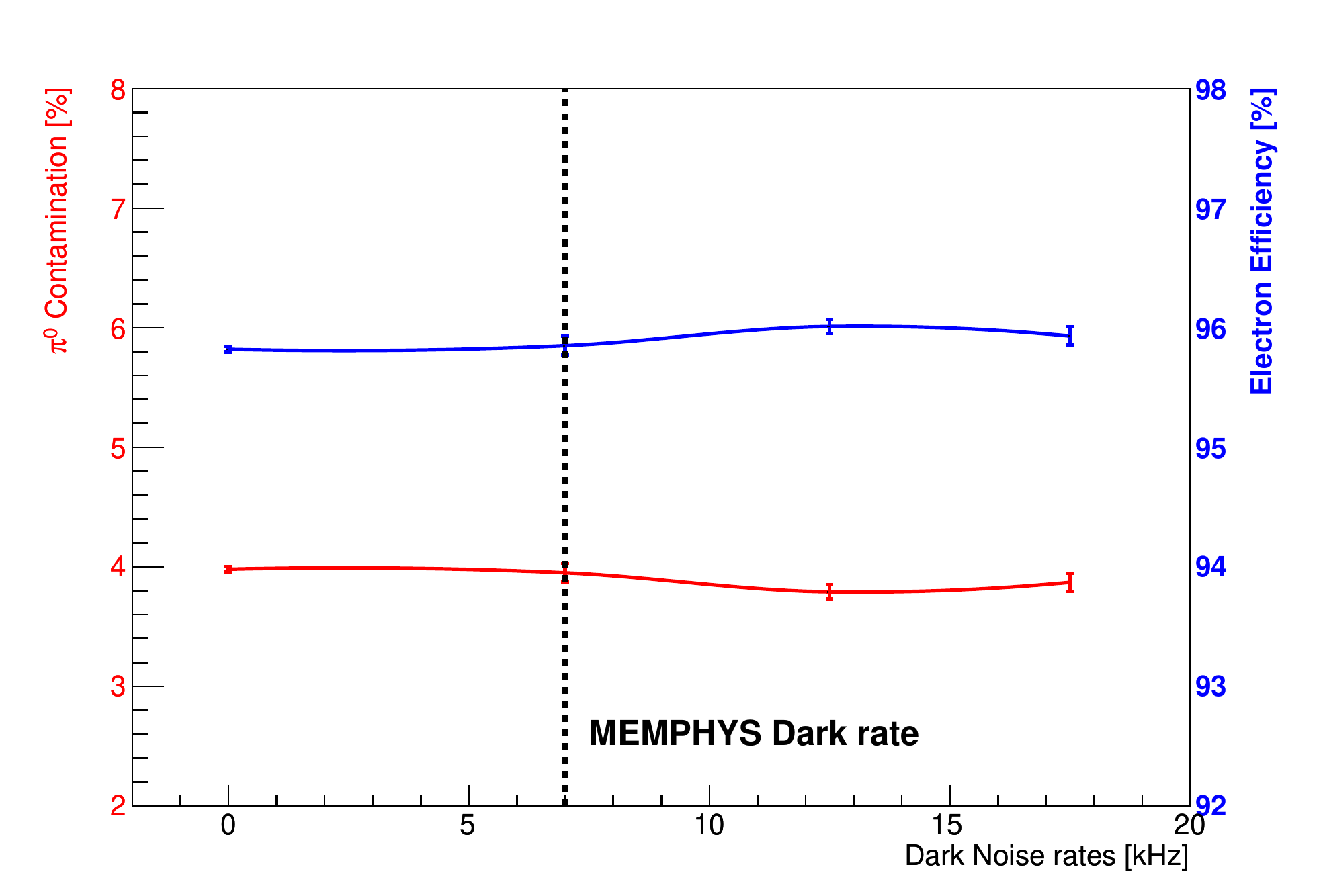} &
\includegraphics[width=0.5\textwidth]{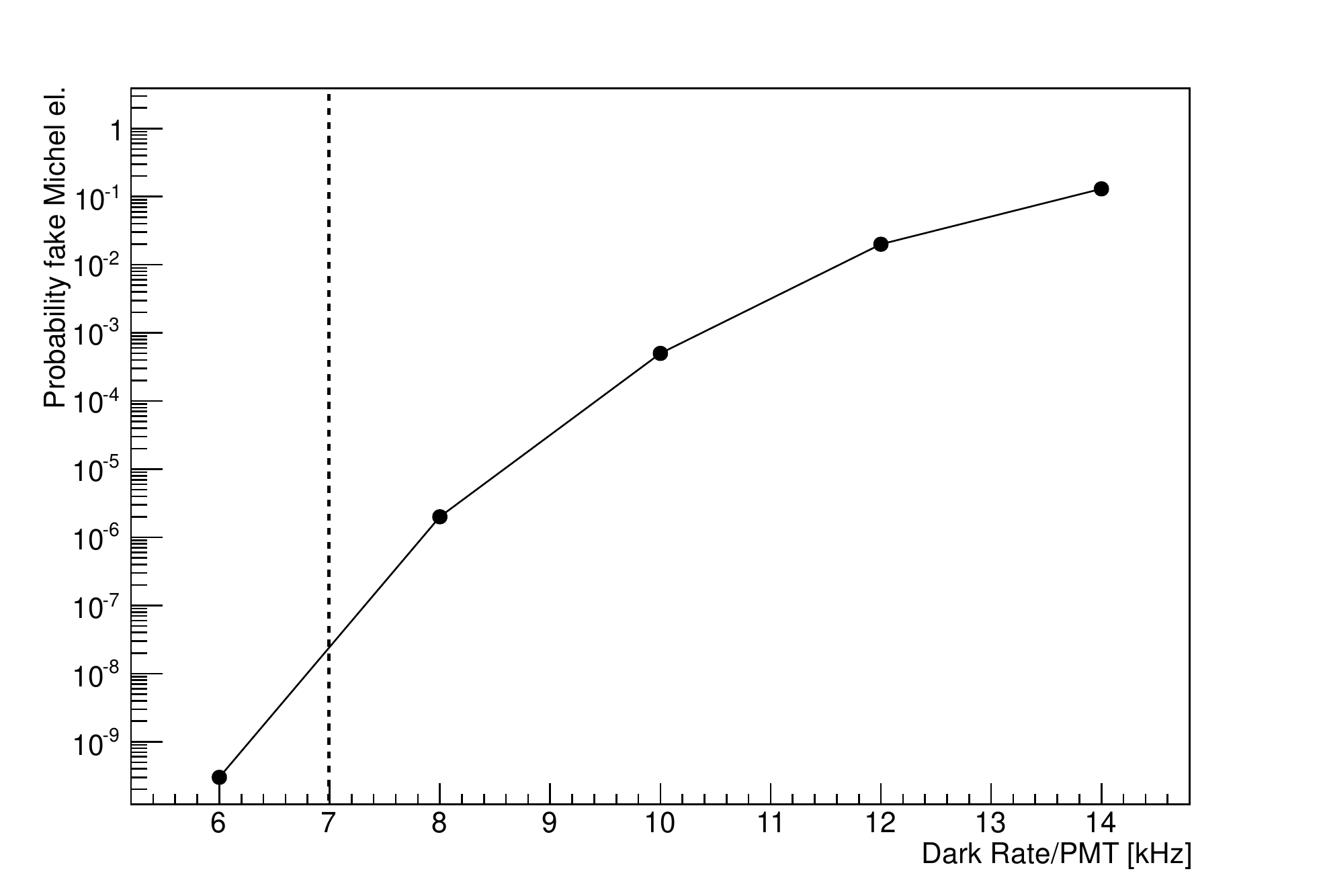}       \\
\end{tabular}
\caption{Study of the impact of PMT dark noise on the analysis.
Left: $\pi^0$ contamination and efficiency for $\nu_e$-CC identification
as a function of dark count rate per PMT. Right: probability of finding a fake
Michel-electron as a function of dark count rate per PMT.
A dark count rate of about 7 kHz per PMT,
shown by the vertical lines, is considered 
as a realistic estimate for the MEMPHYS detector.
\label{fig:dark}
}
}
\end{figure}

PMT dark noise could in principle affect the analysis by worsening the performance of
selection cuts where a small number of hits can have a significant impact.
One example is the selection of $\pi^0$s, where the second ring can be faint.
Another one is the tagging of Michel electrons following muon decays,
as noise hits can mimic a low energy electron after $\nu_e$-CC events and thus reduce the efficiency for
their identification or increment their contribution to the
background of the $\nu_{\mu}$-CC sample.
PMT noise was implemented in the simulation in order to evaluate its impact, 
in particular on these two cuts.
The efficiency for
single-ring identification and the $\pi^0$ contamination in $\nu_e$-CC events were evaluated as 
a function of PMT dark rate, and found to be quite insensitive to it, as shown in 
Figure~\ref{fig:dark} (left). The probability to tag a fake Michel electron is shown as a function 
of dark count rate in Figure~\ref{fig:dark} (right), and is extremely low for
the dark count rate expected in MEMPHYS. We can conclude that
no significant degradation of the analysis performance should be expected from PMT dark noise.

\begin{figure}[t]
\center{
\begin{tabular}{cc}
\includegraphics[width=0.5\textwidth]{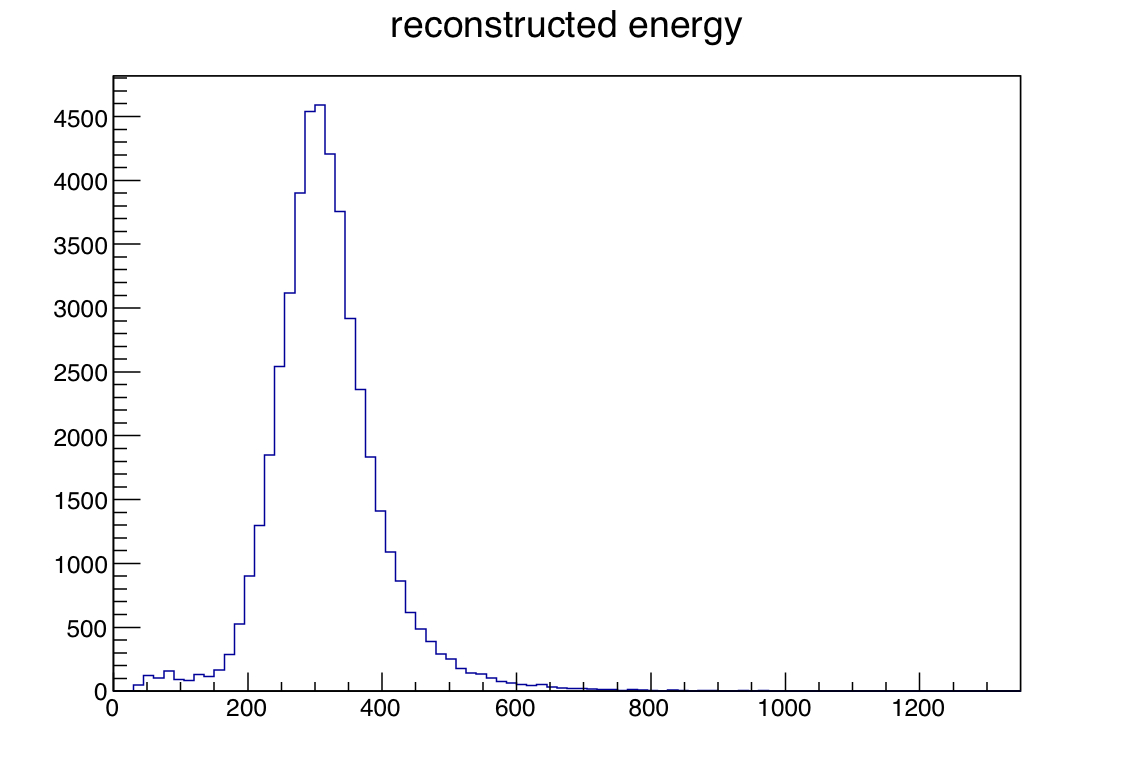} &
\includegraphics[width=0.5\textwidth]{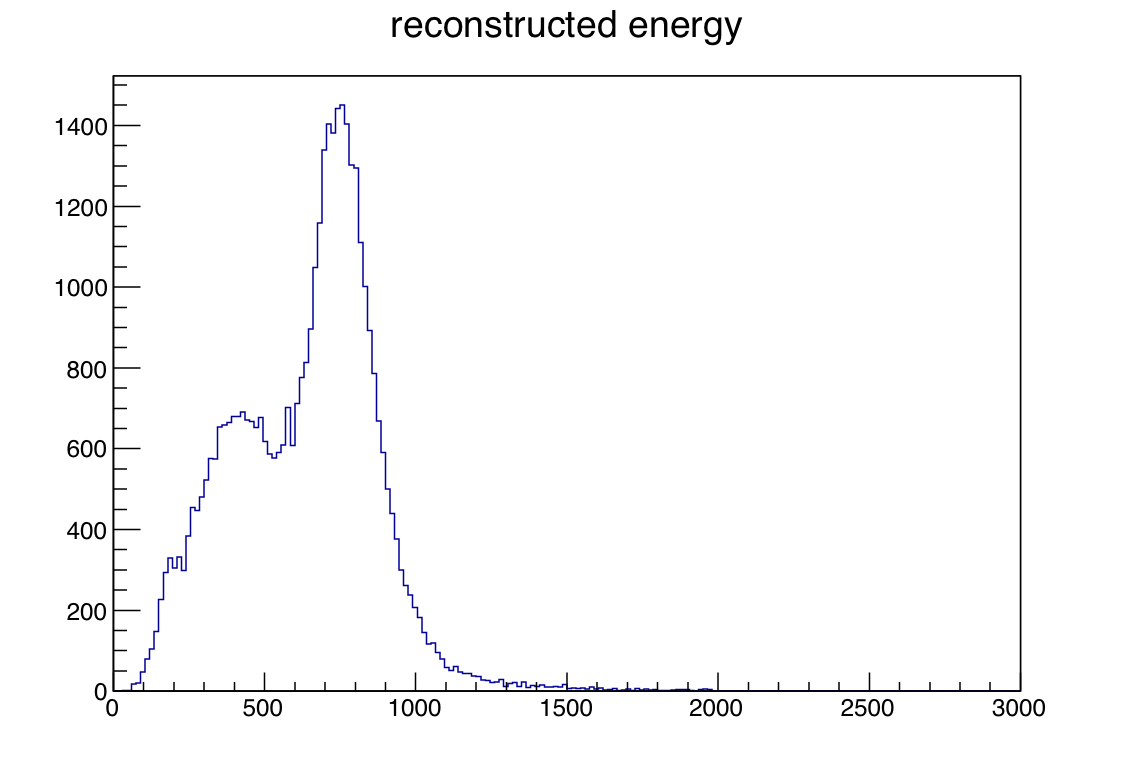}       \\
\vspace{-0.5cm} & \vspace{-0.5cm} \\
\tiny MeV & \tiny MeV
\end{tabular}
\caption{
Reconstructed energy
for selected muon neutrinos with energies of 360 MeV (left) 
and 840 MeV (right). 
\label{fig:enuresol}
}
}
\end{figure}

The incident neutrino energy is deduced from the measured
lepton momentum and direction, assuming the interaction to be CC
and quasi-elastic (QE).
In a pure 2-body collision $\nu_l+N \to l+N'$ (where $l$=$e$ or $\mu$
and $N$ denotes a nucleon, either $p$ or $n$), and assuming the nucleon 
is at rest, the incoming neutrino 
energy $E_\nu$ is related by simple kinematics
to the outgoing lepton energy $E_l$ and momentum $P_l$ and to the angle $\theta_l$
of the lepton direction with respect to the neutrino:

\begin{equation} E_\nu = \frac{m_NE_l-m_l^2/2}{m_N-E_l+P_l\cos\theta_l}
\label{eq:Enu} \end{equation}

The difference between the reconstructed and true neutrino energy 
in two different energy ranges
is shown in figure~\ref{fig:enuresol}: the Gaussian peak is due to true
QE interactions,
with a smearing induced by the Fermi motion of the nucleon and the experimental
resolution, while the tail at lower reconstructed energies
is due to non-QE interactions, whose contribution is larger
as the neutrino energy increases.

\section{Migration matrices}

In order to properly take into account all the effects of the reconstruction,
the detector performance is conventionally described in terms of ``migration matrices''
representing the reconstructed neutrino energy as a function of the true one.
Each ``slice'' of true energy is normalized such that the projection of the matrix
corresponds to the efficiency for the given neutrino energy.
Separate matrices are constructed for signal and background in the different 
detection channels, and for CC and NC events.

Events identified as electron neutrinos are the signal in the appearance
channel in a ``traditional'' neutrino beam (Super-Beam)~\cite{longhin}, 
composed mainly of $\nu_\mu$'s, where the
oscillation $\nu_\mu\to\nu_e$ is searched. 
Separate migration matrices are provided for CC and NC interactions.
The background is given by mis-identified $\nu_\mu$ CC interactions
as well as by other components present in the beam in small fraction 
(mainly $\nu_e$'s and anti-neutrinos; no detailed study has been performed here for
$\nu_\tau$'s, since the beam energy is below the threshold for $\tau$ production). 
Events identified as muons are the signal for the appearance channel 
$\nu_e\to\nu_\mu$ with a Beta-Beam~\cite{zucchelli,bbbook} or for the disappearance
channel $\nu_\mu\to\nu_\mu$ with a Super-Beam.

\begin{figure}[t]
\center{
\begin{tabular}{cc}
{\fbox{$\nu_\mu$ selection}} & {\fbox{$\nu_e$ selection}} \\
\vspace{2pt}  & \vspace{2pt} \\
\includegraphics[width=0.5\textwidth]{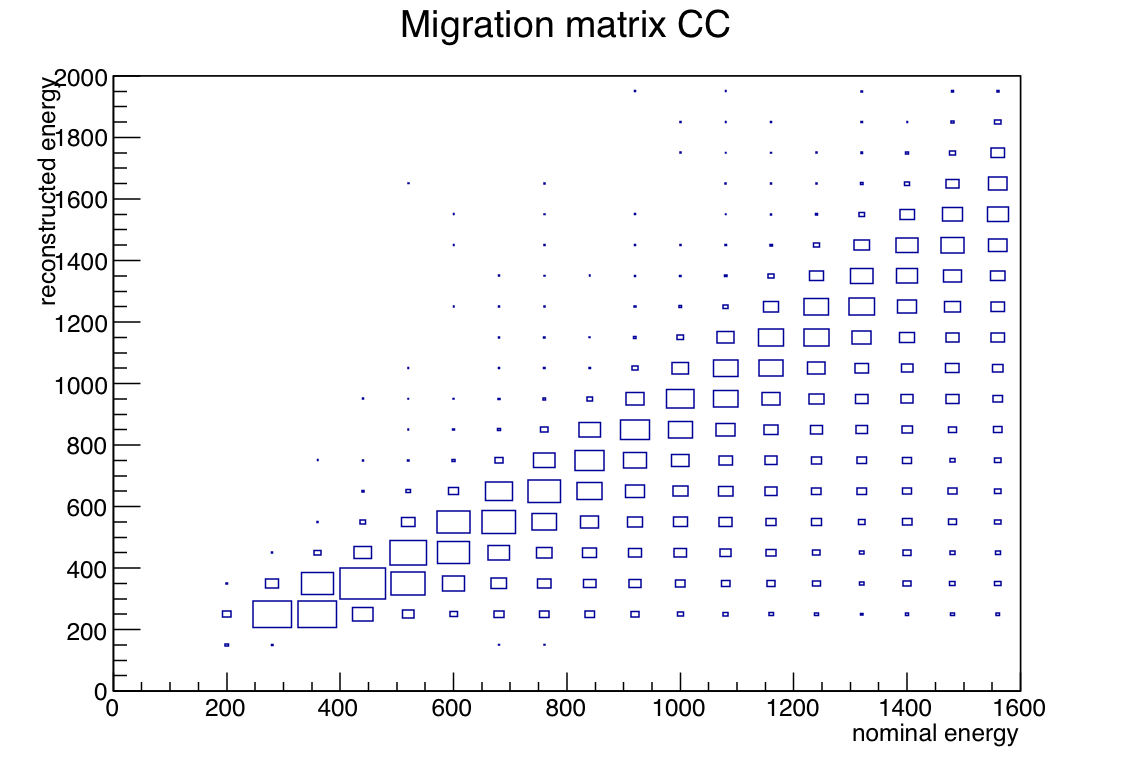} &
\includegraphics[width=0.5\textwidth]{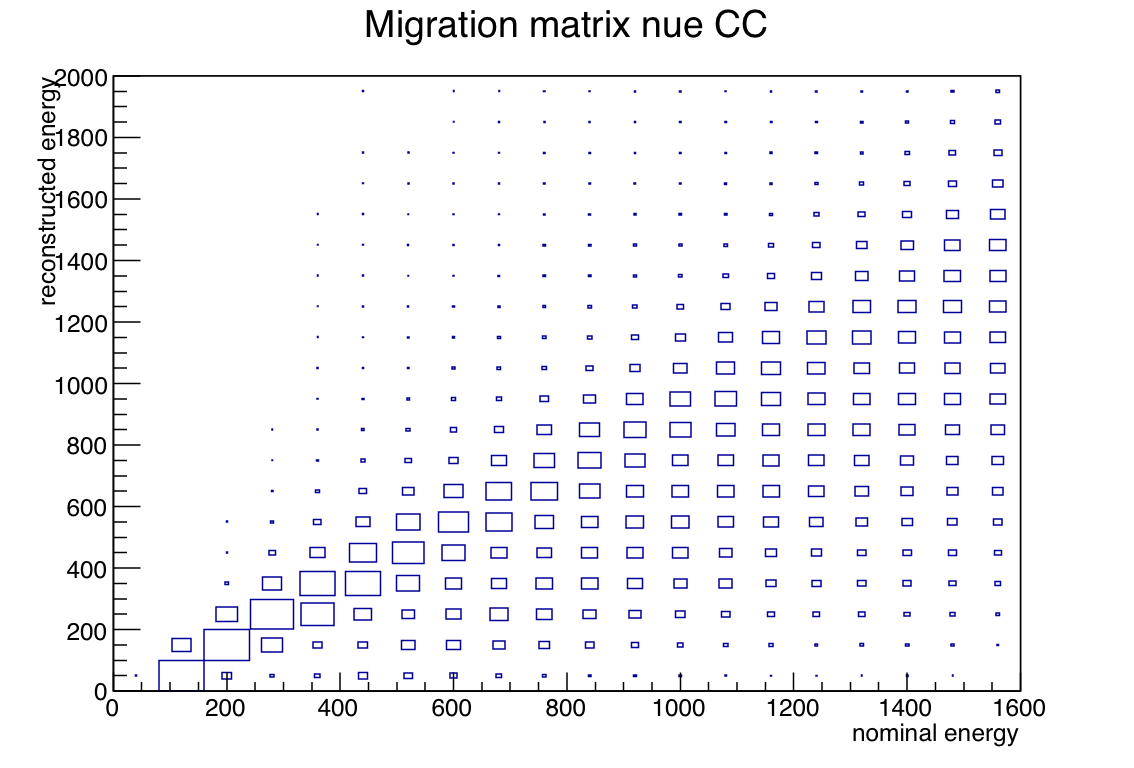} \\
\includegraphics[width=0.5\textwidth]{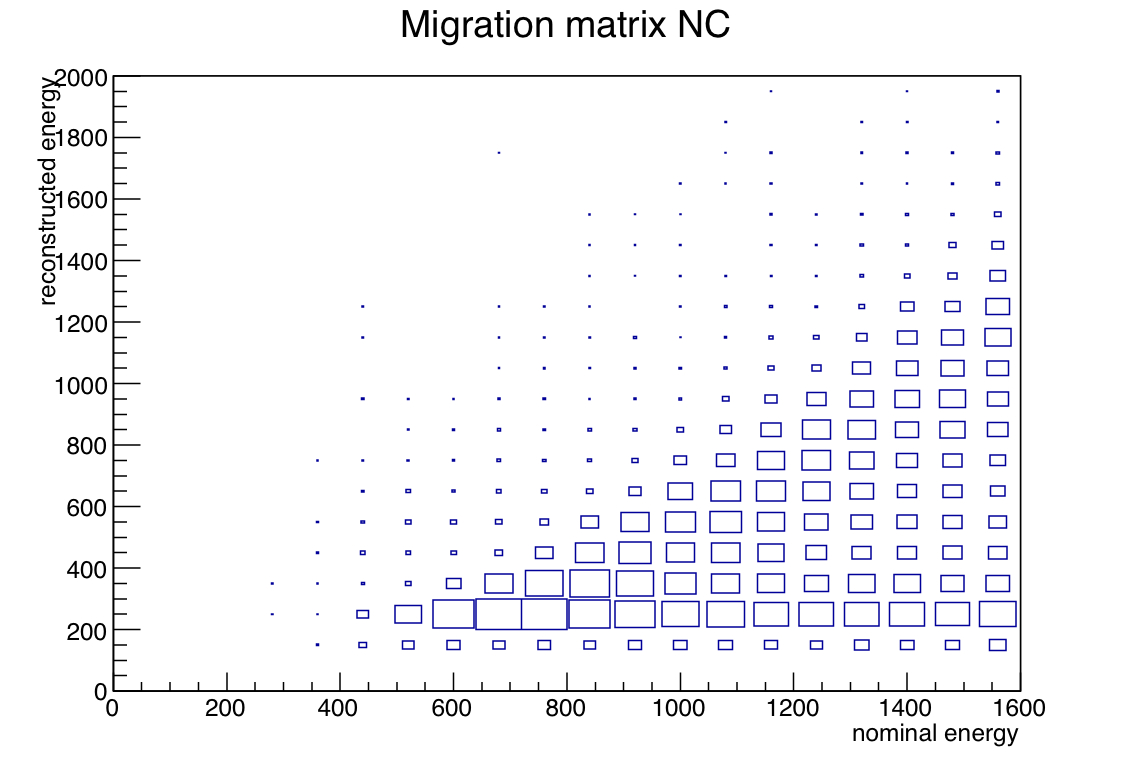} &
\includegraphics[width=0.5\textwidth]{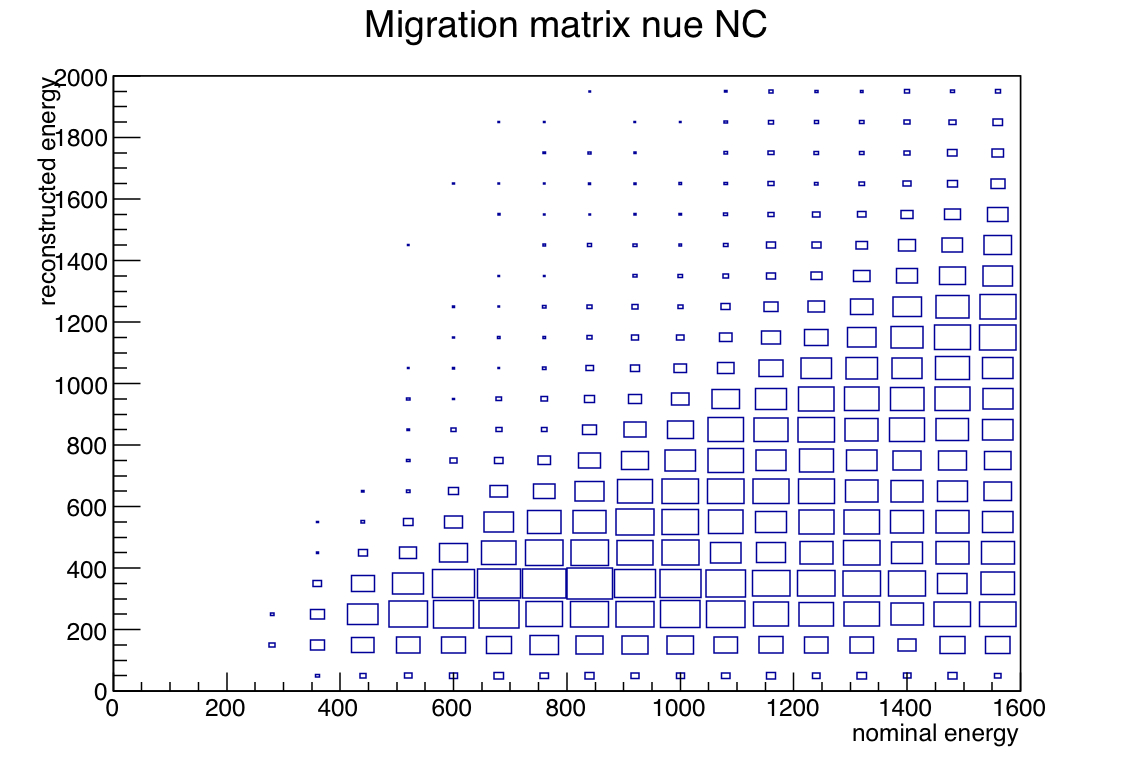} \\
\includegraphics[width=0.5\textwidth]{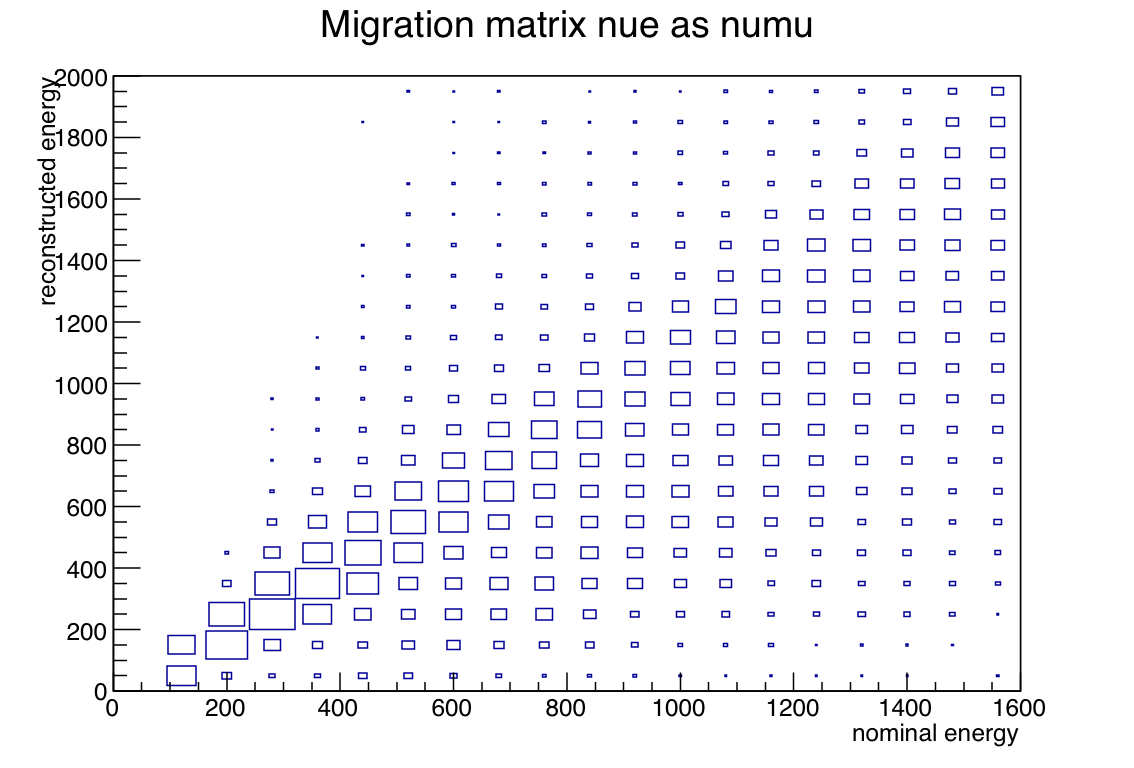} &
\includegraphics[width=0.5\textwidth]{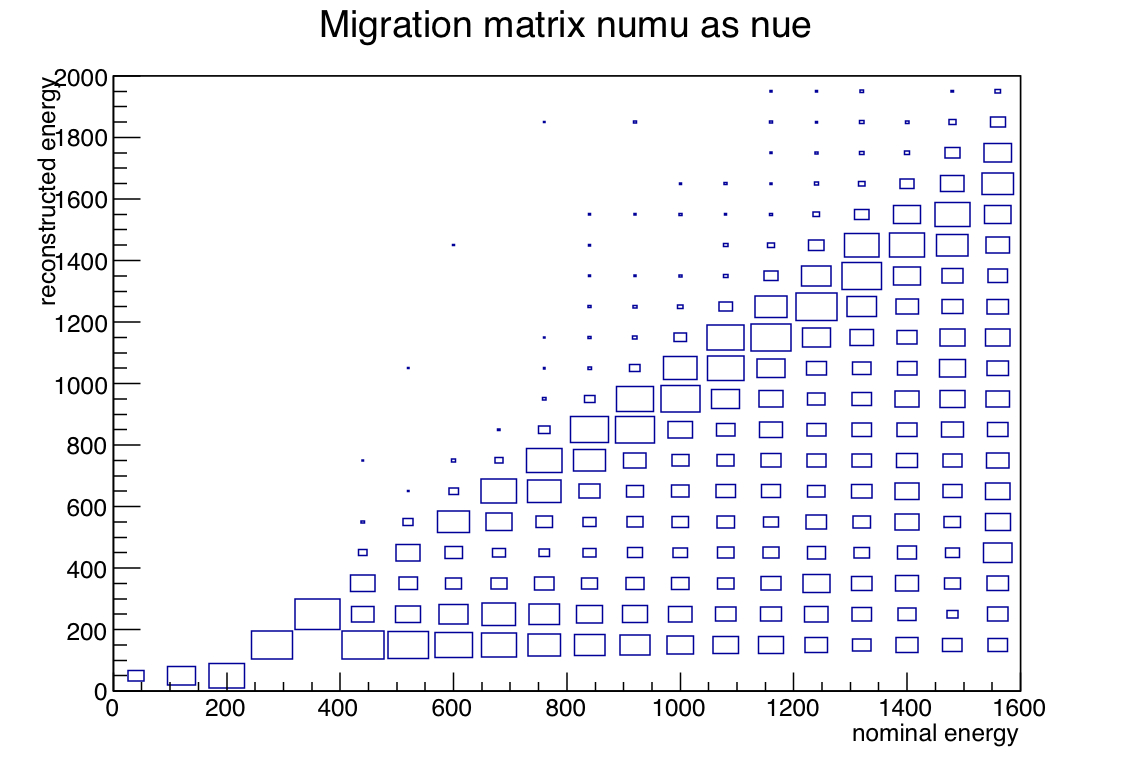} \\
\end{tabular}
\caption{
``Migration matrices'' with reconstructed neutrino energy
as a function of true energy for selected events.
Left: events identified as muon neutrinos,
when they are $\nu_\mu$ CC interactions (top), 
NC interactions (middle), $\nu_e$ CC interactions (bottom).
Right: events identified as electron neutrinos,
when they are $\nu_e$ CC interactions (top), 
NC interactions (middle), $\nu_\mu$ CC interactions (bottom).
\label{fig:migmats}
}
}
\end{figure}

\begin{figure}[t]
\center{
\begin{tabular}{cc}
{\fbox{$\nu_\mu$ selection}} & {\fbox{$\nu_e$ selection}} \\
\vspace{2pt}  & \vspace{2pt} \\
\includegraphics[width=0.5\textwidth]{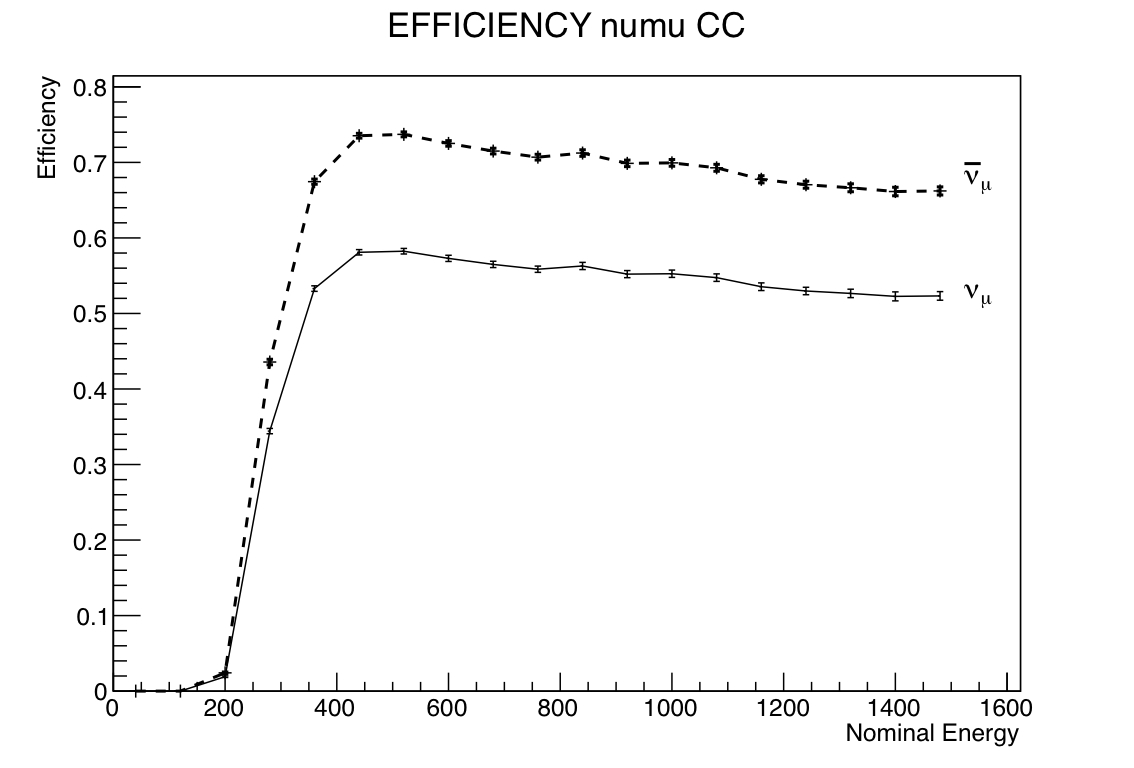} &
\includegraphics[width=0.5\textwidth]{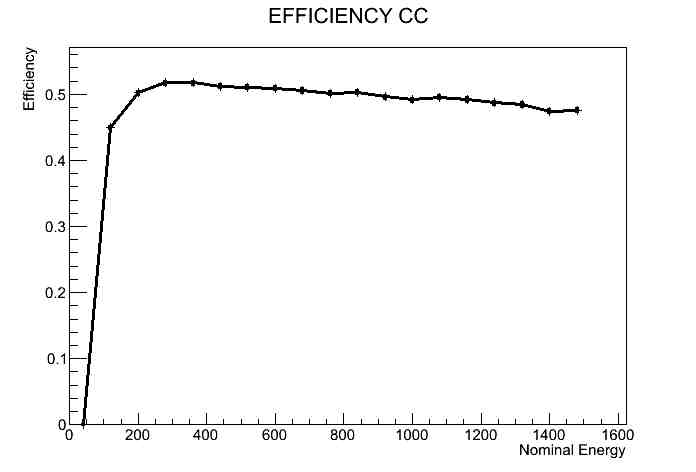} \\
\includegraphics[width=0.5\textwidth]{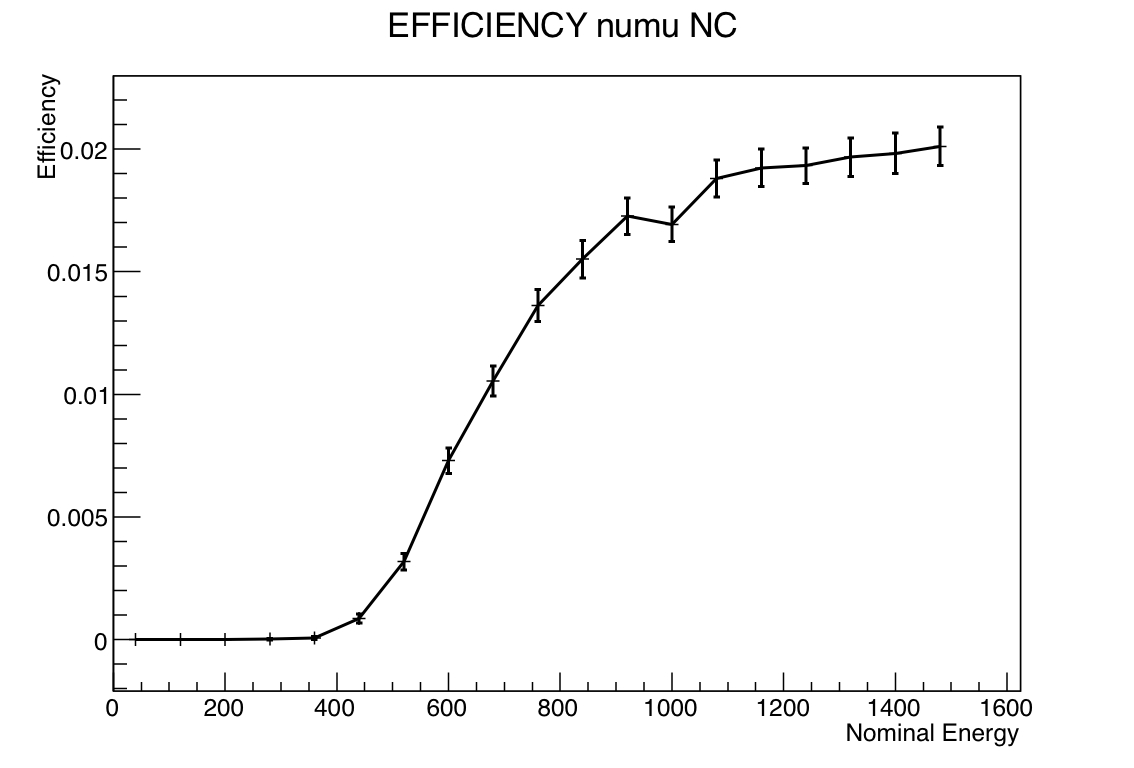} &
\includegraphics[width=0.5\textwidth]{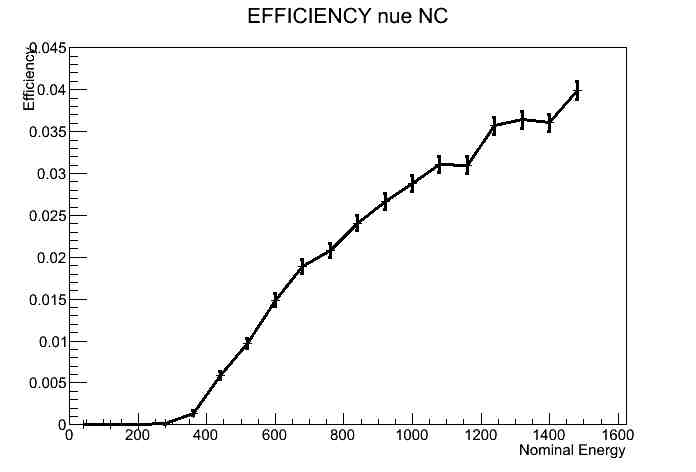} \\
\includegraphics[width=0.5\textwidth]{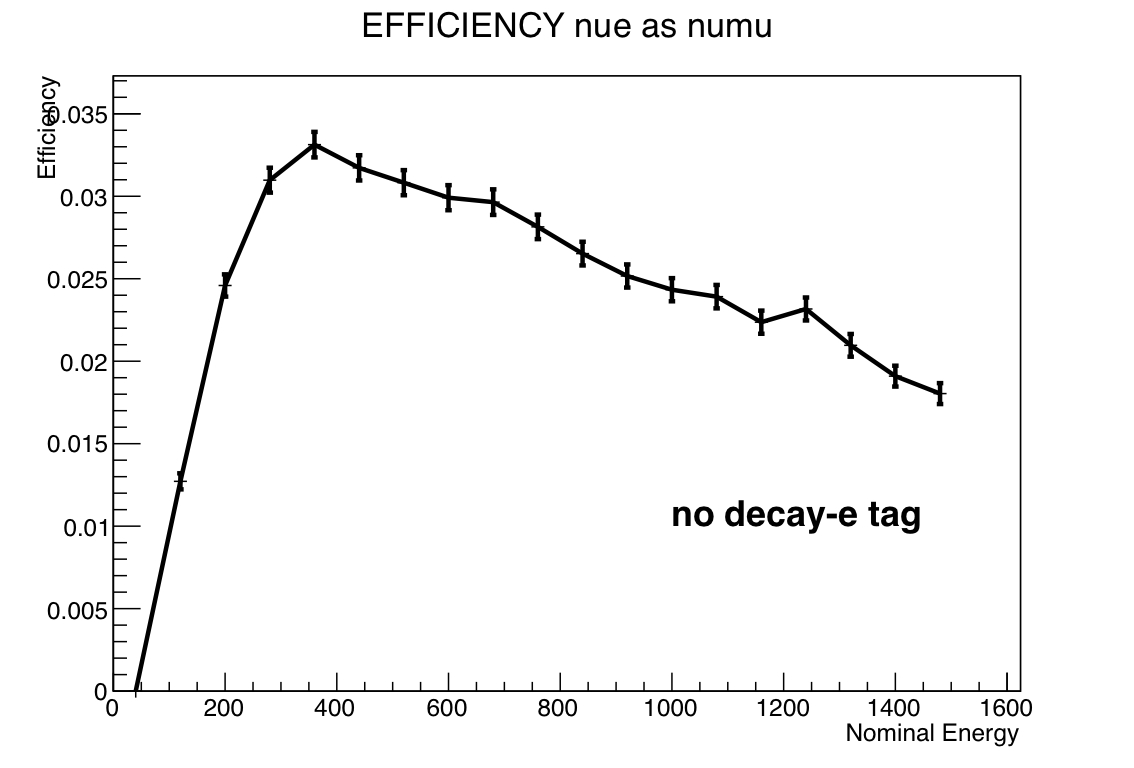} &
\includegraphics[width=0.5\textwidth]{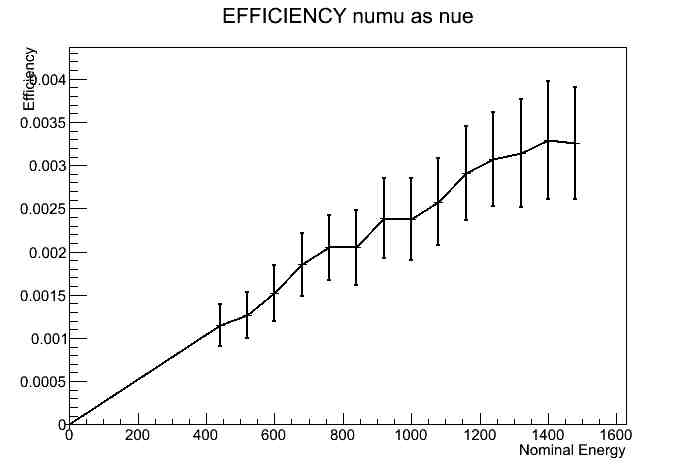} \\
\end{tabular}
\caption{
Efficiencies for the selection of the
different neutrino event categories in the MEMPHYS detector,
as a function of neutrino energy. 
Left: events identified as muon neutrinos,
when they are $\nu_\mu$ CC interactions (top), 
NC interactions (middle), $\nu_e$ CC interactions 
(bottom. The cut decay-electron tag 
completely suppresses 
$\nu_e$ CC interactions and has not been applied for this plot).
Right: events identified as electron neutrinos,
when they are $\nu_e$ CC interactions (top), 
NC interactions (middle), $\nu_\mu$ CC interactions (bottom).
\label{fig:effs}
}
}
\end{figure}

The details of the matrices are provided in figure~\ref{fig:migmats}.
The efficiencies as a function of neutrino energy
are shown in figure~\ref{fig:effs}.

\begin{figure}[t]
\center{
\begin{tabular}{cc}
\includegraphics[width=0.5\textwidth]{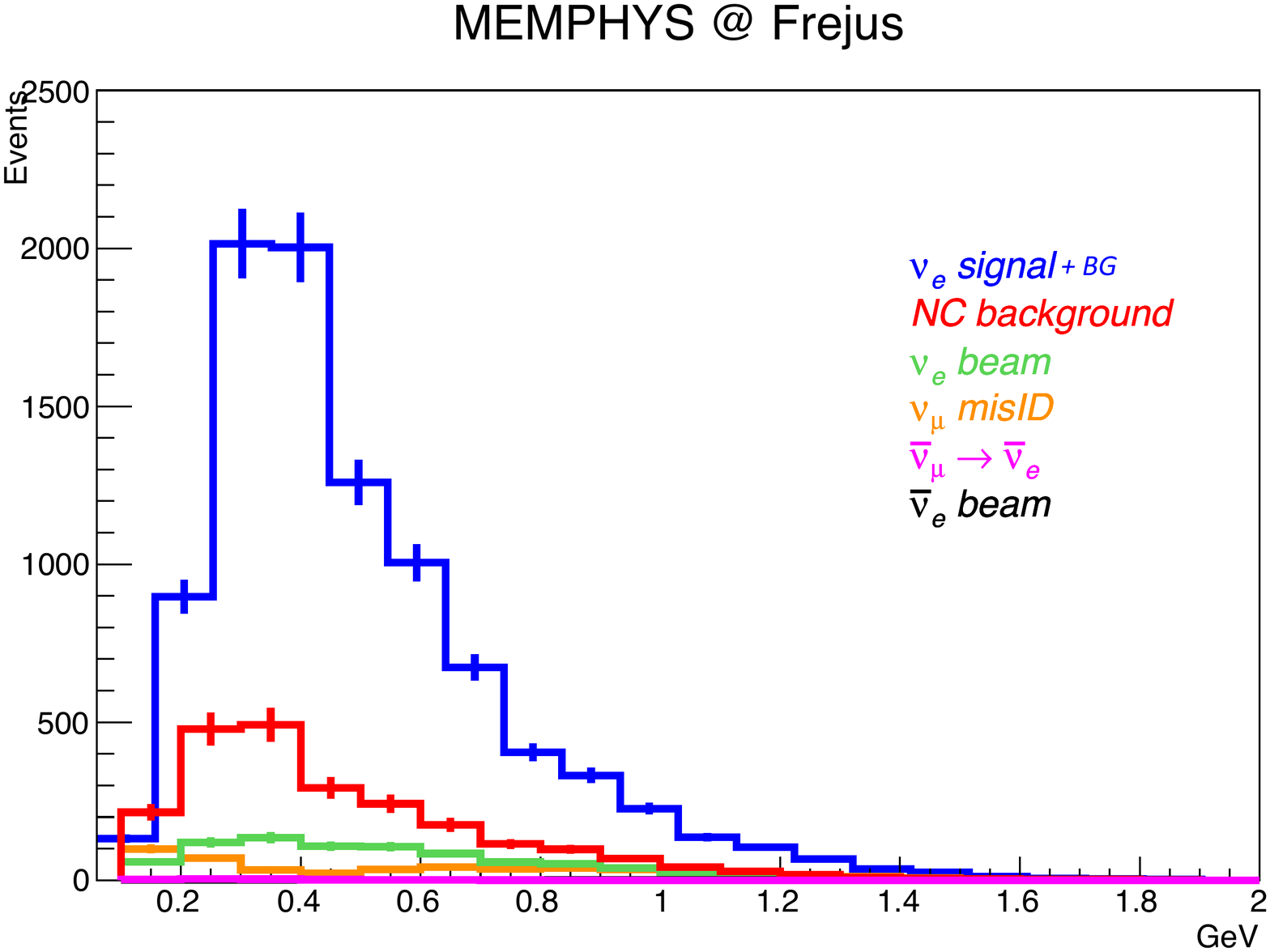} &
\includegraphics[width=0.5\textwidth]{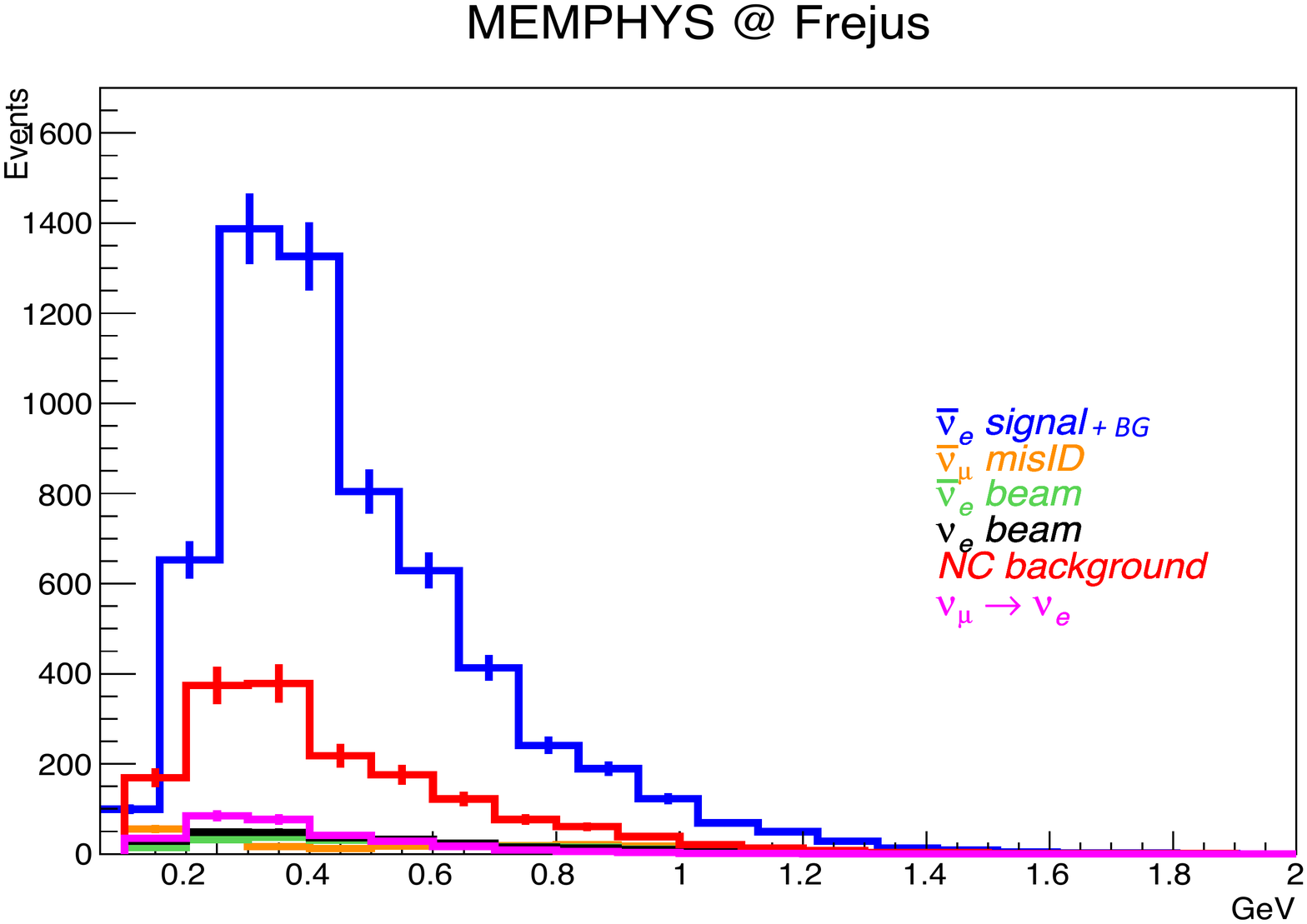} \\
\end{tabular}
\caption{
Neutrino (left) and antineutrino (right) spectra measured in MEMPHYS with
a Super-Beams from CERN to the Fr\'ejus site, obtained with our migration matrices.
The running time is 2 years and 8 years respectively.
$\sin^22\theta_{13}=0.1$ and $\delta_{CP}=0$ are assumed.
\label{fig:spectra}
}
}
\end{figure}

Examples of neutrino and antineutrino spectra measured in the MEMPHYS detector 
are shown in Figure~\ref{fig:spectra}.
They are obtained
with the Super-Beam fluxes provided by A.Longhin~\cite{longhin},
from CERN to the Fr\'ejus site,
and using the migration matrices to account for experimental effects.

The matrices are available from the authors
in the text format suitable as input for the GLOBES package~\cite{globes-1,globes-2}.

\begin{figure}[t]
\center
{\begin{tabular}{cc}
\includegraphics[width=0.5\textwidth]{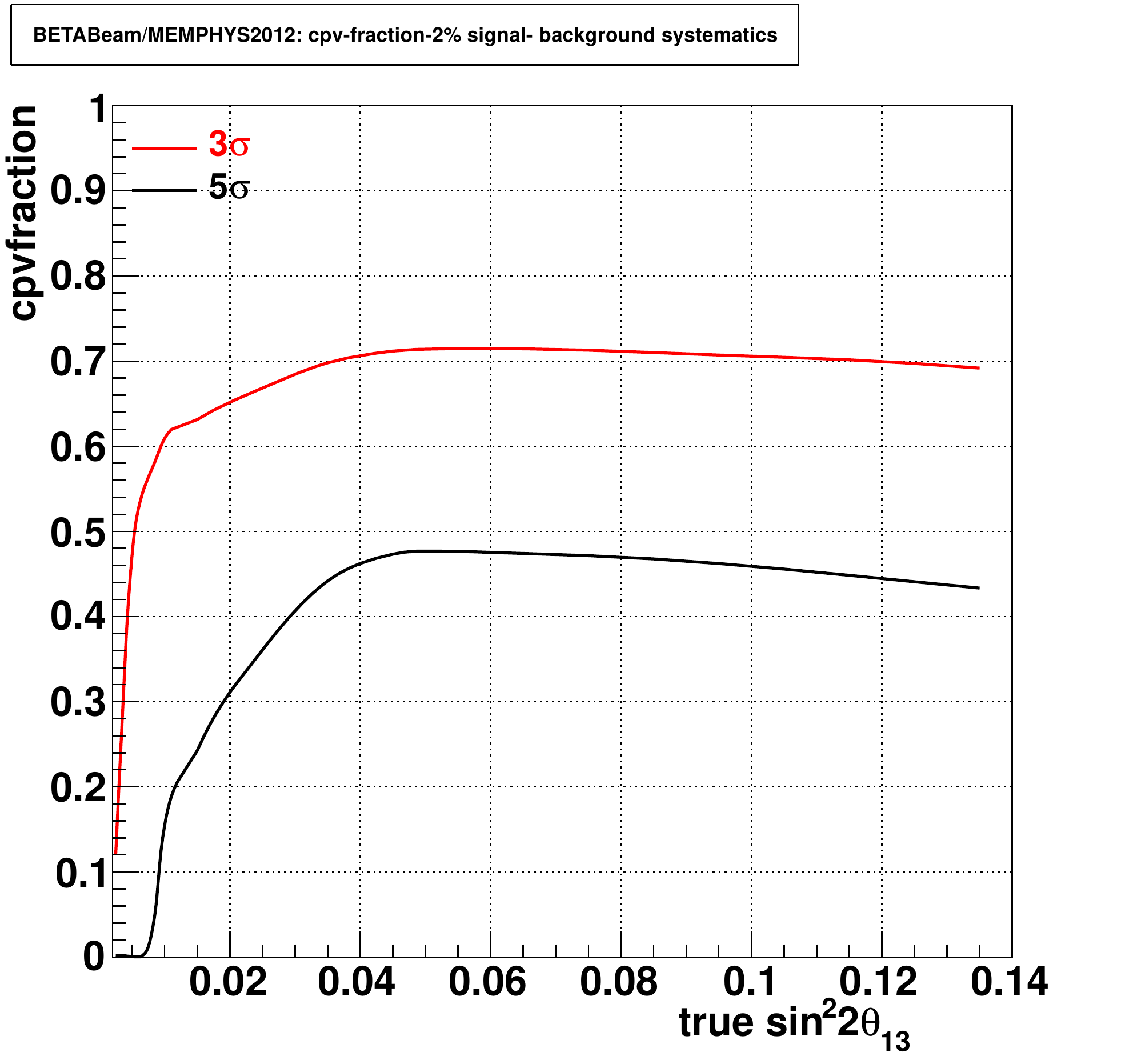} &
\includegraphics[width=0.5\textwidth]{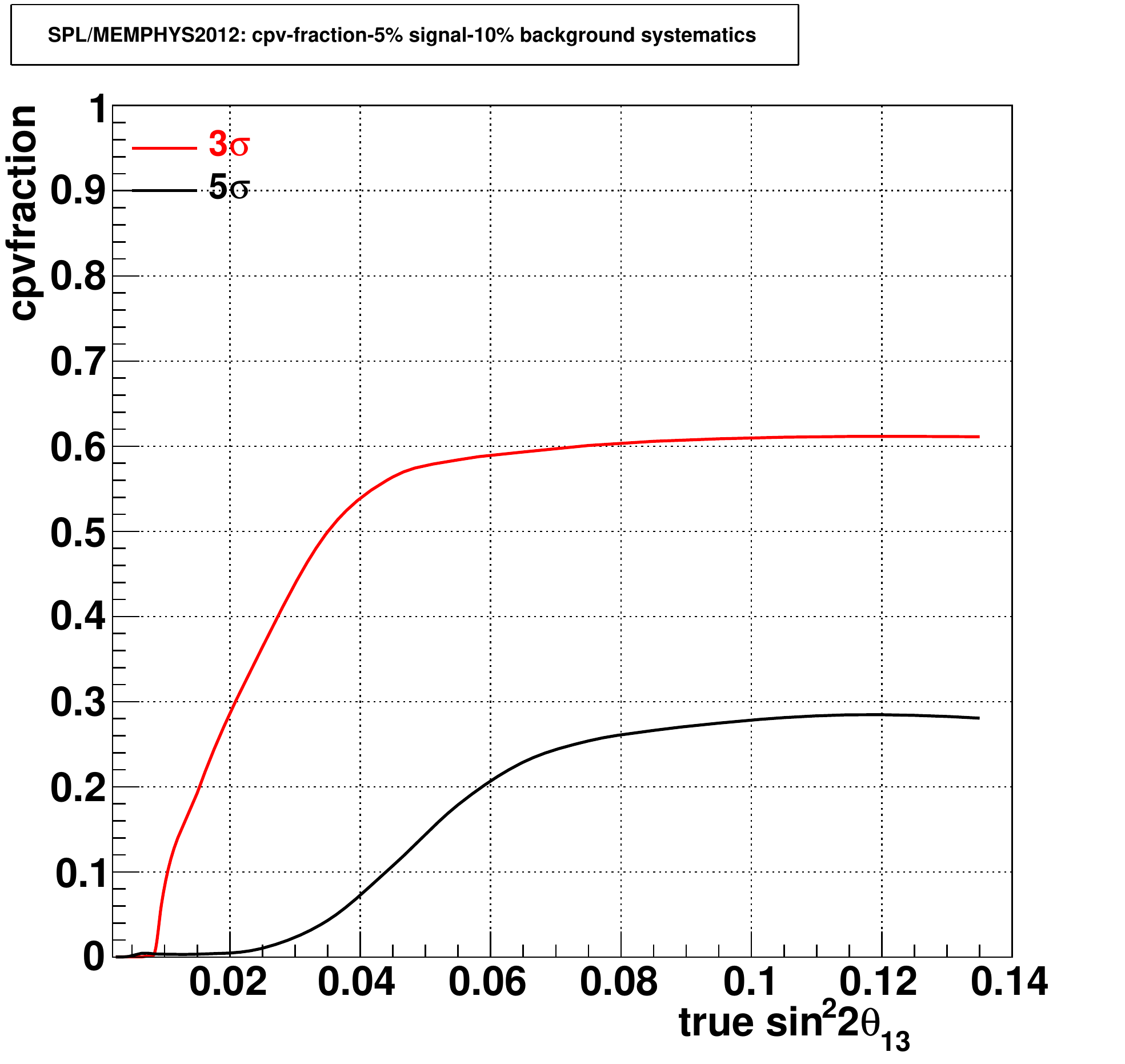}
\end{tabular}
}
\caption{
Example of study of sensitivity to the 
leptonic CP violation phase using the GLOBES package,
considering a Beta-Beam (left) 
or a Super-Beam (right) from CERN to the Fr\'ejus site.
\label{fig:globes}
}
\end{figure}

Figure~\ref{fig:globes} shows an example of study of sensitivity to the 
leptonic CP violation phase using the GLOBES package,
with a Beta-Beam~\cite{jec} 
and a Super-Beam~\cite{longhin} from CERN to Fr\'ejus.
For the Beta-Beam, 
a running time of 5 years with neutrinos and 5 years with antineutrinos is considered,
with a systematic uncertainty of 2\% on both signal and background.
For the Super-Beam, 
a running time of 2 years with neutrinos and 8 years with antineutrinos is considered,
with a systematic uncertainty of 5\% on signal and 10\% on background.
Normal mass hierarchy is assumed.
The sensitivity to the CP violation phase in the leptonic sector, $\delta_{CP}$,
is shown, at 3$\sigma$ and 5$\sigma$, as a function of the $\theta_{13}$ mixing angle.

\section{Conclusions}

A detailed study of the performance of a future large-scale water-Cherenkov detector,
MEMPHYS, has been performed, using a full simulation of the detector's response
and realistic analysis algorithms.
The results have been presented in terms of migration matrices
from true to reconstructed neutrino energy, considering the signal and background
channels for different neutrino beam types.

\acknowledgments
We are grateful to Enrique Fernandez Martinez, Pilar Coloma and Sanjib Agarwalla
for useful discussion.
We acknowledge the financial support of the European Community under the
European Commission Framework Programme 7 Design Study EUROnu, Project
Number 212372 and Design Study LAGUNA-LBNO, Project Number 284518. 
The EC is not liable for any use that may be made of the
information contained herein.

\bibliography{MMshort}

\end{document}